\documentclass[12pt]{article}
\usepackage{amsmath, amssymb}
\usepackage{graphicx}
\usepackage{pstcol,epic,epsfig}
\begin{document}
\date{}

\title{SUSY QM, symmetries and spectrum generating algebras for two-dimensional systems}

\author{D. Mart\'{\i}nez$^{a}$\footnote{{\it E-mail address:} dmartinezs77@yahoo.com.mx} and R. D. Mota$^{b}$ } \maketitle

\begin{minipage}{0.9\textwidth}
\small $^{a}$Escuela Superior de F\'{\i}sica y Matem\'aticas,
Instituto Polit\'ecnico Nacional,
Ed. 9, Unidad Profesional Adolfo L\'opez Mateos, 07738 M\'exico D F, M\'exico.\\
$^{b}$Unidad Profesional Interdisciplinaria de Ingenier\'{\i}a y
Tecnolog\'{\i}as Avanzadas, IPN. Av. Instituto Polit\'ecnico
Nacional 2580, Col. La Laguna Ticom\'an, Delegaci\'on Gustavo A.
Madero, 07340 Mexico D. F., Mexico.
\end{minipage}
\rm
\begin{abstract}
We show in a systematic and clear way how factorization methods can
be used to construct the generators for hidden and dynamical symmetries.
This is shown by studying the $2D$ problems of hydrogen atom,
the isotropic harmonic oscillator and the radial potential
$A\rho^{2\zeta-2}-B\rho^{\zeta-2}$. We show that in these cases the
non-compact (compact) algebra corresponds to  $so(2,1)$ ($su(2)$).\\
\end{abstract}

\section{Introduction}

Since their introduction the factorization methods have played an
important roll in the study of quantum systems
\cite{DIRAC,SCHRO1,SCHRO2}. This is because, if the Schr\"odinger
equation is factorizable, the energy spectrum and the eigenfunctions
are obtained algebraically. Infeld and Hull \cite{INFELD},
generalized the ideas of Dirac \cite{DIRAC} and Schr\"odinger
\cite{SCHRO1,SCHRO2} and created a new factorization method (IHFM)
which uses a particular solution for the Ricatti equation. This
method allowed to classify the problems according to the
characteristics involved in the potentials. A variant of the IHFM,
proposed by Mielnik \cite{MIELNIK}, gives the possibility to obtain
isospectral Hamiltonians. On the other hand, the charges in $N=2$
SUSY QM are written in terms of the operators which factorize the partner Hamiltonians  $H_+$ and $H_-$
\cite{JUNKER, BAGCHI,COOPER01}.

For the most important central potential problems (the hydrogen atom
and the isotropic harmonic oscillator) algebraic treatments by means
of non-compact and compact groups are known
\cite{WYBOURNE,ENGLEFIELD}, being the last ones generally related to
hidden symmetries and degeneracy \cite{FOCK}-\cite{MOSHI2}. However,
a systematic method to find both compact and non-compact group
generators for a given system has not still developed. These
generators have been intuitivelly found and forced them to close an algebra, as it is extensively shown in reference
\cite{BOHM}.

The link between SUSY QM and hidden symmetries has been studied  by
showing that the factorization operators of the partner Hamiltonians
are contained within the symmetry group operators
\cite{LYM}-\cite{IOFFE}. This connection has been achieved for the
two-  and three-dimensional hydrogen atom \cite{LYM, MOTA1}, the
two- and three-dimensional isotropic harmonic oscillator
\cite{MOTA2, MOTA3} and for a neutron in the magnetic field of a
linear current \cite{DAN1}-\cite{IOFFE}. It must be emphasized
that the charges are written in terms of the superpotential which is
involved in the factorization operators of the Hamiltonian $H_+$.
Thus, the symmetries of the Hamiltonian $H_+$ play an important role
in the construction of operators which define a superalgebra of the
SUSY Hamiltonian $H=$ diag $(H_+,H_-)$. Also, hidden supersymmetry,
related with additional supersymmetries specific to the dynamics,
has been associated with the existence of a supergroup ( a dynamical
superalgebra ) \cite{VINET1}-\cite{MACFAR}.

The purpose of this paper is to show that the IHFM allows us to
construct the generators of symmetries (compact algebras), whereas
the Schr\"odinger factorization allows us to find dynamical
symmetries (spectrum generating algebras).

This work is organized as follows. In Sections 2 and 3 we study,
from the factorization methods approach, the two-dimensional
isotropic harmonic oscillator (TDIHO) and the two-dimensional
hydrogen atom (TDHA), respectively. By means of the Schr\"odinger
factorization we obtain in a systematic way the corresponding
non-compact algebra generators and show that these systems have an
$so(2,1)$ symmetry. On the other hand, for each problem we study how
the constants of motion and the radial SUSY operators are related.
In section 4 we use the results from the previous sections to find
the dynamical and hidden symmetries for the radial potential
$A\rho^{2\zeta-2}-B\rho^{\zeta-2}$, which generalizes the former
ones \cite{KHARE1,KHARE2,VINET}. Finally, in Section 5 we give the
concluding remarks.

\section{Two-dimensional isotropic harmonic oscillator}
\subsection{Schr\"odinger factorization}
The Schr\"odinger equation for the two-dimensional isotropic
harmonic oscillator (TDIHO) in polar coordinates $(\rho,\varphi)$ is
\begin{equation}
H_{h.o.}\psi(\rho,\varphi)\equiv\left[-\frac{\hbar^2}{2\mu}\left(\frac{\partial^2}{\partial\rho^2}+\frac{1}{\rho}\frac{\partial}{\partial\rho}+
\frac{1}{\rho^2}\frac{\partial^2}{\partial\varphi^2}\right)+\frac{\hbar\omega}{2}\beta^2\rho^2\right]\psi(\rho,\varphi)=E\psi(\rho,\varphi),
\label{Hoab}
\end{equation}
where $\beta=\sqrt{\mu\omega/\hbar}$. In what follows, we will set $\hbar=\omega=1$ and assume that the mass involved in each system under study is $\mu=1$.

Considering the conservation of the operator $L_z$, the
eigenfunction can be written as
\begin{equation}
\psi_{nm}(\rho,\varphi)=R_{nm}(\rho)e^{im\varphi}\equiv\rho^{-\frac{1}{2}}u_{nm}(\rho)e^{im\varphi},
\label{psioab}
\end{equation}
where $m=0,\pm1,\pm2,...$ is the quantum number corresponding to the
operator $L_z$. Therefore, from equation (\ref{Hoab}) we obtain
\begin{equation}
H_mu_{nm}\equiv-\frac{1}{2}\frac{d^2u}{d\rho^2}+\left[
\frac{1}{2}\rho^2+\frac{1}{2}\frac{m^2-\frac{1}{4}}{\rho^2}\right]u_{nm}=E_n
u_{nm}. \label{habrh}
\end{equation}
This equation can be written as
\begin{equation}
\left(\rho^2\frac{d^2}{d\rho^2}-\rho^4+2E_n
\rho^2\right)u_{nm}=\left(m^2-\frac{1}{4}\right)u_{nm}.
 \label{Hscho}
\end{equation}
We propose a pair of first order operators such that
\begin{equation}
\left(\rho\frac{d}{d\rho}+a\rho^2+b\right)\left(\rho\frac{d}{d\rho}+c\rho^2+f\right)u_{nm}=gu_{nm},
\label{sch}
\end{equation}
where $a,\:b,\:c,\:f$ and $g$ are constants to be determined. By
expanding this expression and comparing it with Eq. (\ref{Hscho}),
we obtain
\begin{equation}
a=-c=\mp1,\hspace{2ex} b=-f=-1\pm E_n,\hspace{2ex} g=m^2-f^2.
\end{equation}
Using these results, Eq. (\ref{sch}) can be expressed in the
following way
\begin{eqnarray}
(D_-^n-1)D_+^nu_{nm}=\frac{1}{4}\left[\left(E_n+\frac{1}{2}\right)\left(E_n+\frac{3}{2}\right)-\left(m^2-\frac{1}{4}\right)\right]u_{nm},\label{dn1}\\
(D_+^n+1)D_-^nu_{nm}=\frac{1}{4}\left[\left(E_n-\frac{1}{2}\right)\left(E_n-\frac{3}{2}\right)-\left(m^2-\frac{1}{4}\right)\right]u_{nm}\label{dn2},
\end{eqnarray}
where we have defined
\begin{equation}
D_{\pm}^n=\frac{1}{2}\left(\mp
\rho\frac{d}{d\rho}+\rho^2-E_n\mp\frac{1}{2}\right). \label{Dn}
\end{equation}
If we consider that the energy spectrum of the TDIHO is
$E_n=n+1$ then, from (\ref{Dn})
the following recursion relations hold
\begin{eqnarray}
D_+^{n\pm2}=D_+^n\mp1,\\
D_-^{n\pm2}=D_-^n\mp1.
\end{eqnarray}
By performing  the change $n\rightarrow n-2$ in Eq. (\ref{dn1}) and
$n\rightarrow n+2$ in Eq. (\ref{dn2}),  we have
\begin{align}
(D_-^{n-2}-1)D_+^{n-2}u_{n-2\;m}& =D_-^n(D_+^n+1)u_{n-2\;m}\nonumber \\
&
=\frac{1}{4}\left[\left(E_n-\frac{1}{2}\right)\left(E_n-\frac{3}{2}\right)-\left(m^2-\frac{1}{4}\right)\right]u_{n-2\;m},\label{dn1b}
\end{align}
and
\begin{align}
(D_+^{n+2}+1)D_-^{n+2}u_{n+2\;m}& =D_+^n(D_-^n-1)u_{n+2\;m}\nonumber \\
&
=\frac{1}{4}\left[\left(E_n-\frac{1}{2}\right)\left(E_n+\frac{3}{2}\right)-\left(m^2-\frac{1}{4}\right)\right]u_{n+2\;m}\label{dn2a},
\end{align}
respectively. Multiplying (\ref{dn1}) by $D_+^n$  and (\ref{dn2}) by
$D_-^n$, and comparing the resulting expressions with (\ref{dn1b})
and (\ref{dn2a}), it is shown that
\begin{align}
D_+^nu_{nm}& \propto u_{n+2\;m},\\
D_-^nu_{nm}& \propto u_{n-2\;m}.
\end{align}
The action of these operators on the radial function $ u_{nm}$ is
shown in figure 1. To find the dynamical symmetry algebra, we define the operator
\begin{equation}
D_3=\frac{1}{4}\left(-\frac{d^2}{d\rho^2}+\rho^2+\frac{m^2-\frac{1}{4}}{\rho^2}\right),
\label{D3}
\end{equation}
which satisfies
\begin{equation}
D_3u_{nm}=\frac{E_n}{2}u_{nm} \label{D3a}.
\end{equation}
Thus, it allows us to define the new pair of operators
\begin{equation}
D_{\pm}=\frac{1}{2}\left(\mp
\rho\frac{d}{d\rho}+\rho^2-2D_3\mp\frac{1}{2}\right), \label{D+-}
\end{equation}
which are independent of $E_n$.

We show that the operators $D_\pm$ and $D_3$ satisfy the $so(2,1)$
Lie algebra
\begin{align}
[D_\pm,D_3]& =\mp D_\pm,\\
\lbrack D_+,D_-\rbrack& =-2D_3.
\end{align}

Thus, from the Schr\"odinger factorization we have obtained the
generators of the non-compact algebra for the TDIHO. In Section 3, a
similar procedure will be applied to find the spectrum generating
algebra for the TDHA.

\subsection{SUSY QM}
Applying SUSY QM to the Hamiltonian $H_m$ defined in Eq.
(\ref{habrh}), we find that the SUSY operators are
\begin{eqnarray}
B_{1,2}=\frac{1}{\sqrt{2}}\Big(\frac{d}{d\rho}\pm\frac{m\pm\frac{1}{2}}{\rho}-\rho\Big),\label{A}\\
B_{1,2}^\dag=\frac{1}{\sqrt{2}}\Big(-\frac{d}{d\rho}\pm\frac{m\pm\frac{1}{2}}{\rho}-\rho\Big)\label{Adag}.
\end{eqnarray}
From now on,  the indices $1$ and $2$ will correspond to the upper
and the lower sign in the expression where they appear. Therefore,
we construct the partner Hamiltonians
\begin{align}
H_{1,2\;+}& \equiv B_{1,2}B_{1,2}^\dag=H_m\mp(m\pm1),\label{Ho+}\\
H_{1,2\;-}& \equiv B_{1,2}^\dag B_{1,2}=H_{m\pm1}\mp m.
\label{Ho-}
\end{align}
By using Eqs. (\ref{Ho+}) and (\ref{Ho-}), we show that
\begin{align}
B_{1}^\dag u_{n\;m+1}& \propto u_{n+1\;m},\\
B_{1}u_{nm}& \propto u_{n-1\;m+1}
\end{align}
and
\begin{align}
B_{2}^\dag u_{n\;m-1}& \propto u_{n+1\;m},\\
B_{2}u_{nm}&\propto u_{n-1\;m-1}.
\end{align}
Thus, the actions of the SUSY operators of the TDIHO on the eigenstates
$u_{nm}$ of equation (\ref{habrh}) are to change the values of both
quantum numbers simultaneously, and they are different among them.
On the other hand, because these operators change the principal
quantum number, they are not directly related to the constants of
motion of the system, as it will be shown in the next section.

\subsection{SUSY QM and constants of motion}
It is well known that the operators
\begin{align}
a_{1,2}& =\frac{1}{2}e^{\pm
i\varphi}\left(\frac{\partial}{\partial\rho}\pm\frac{i}{\rho}\frac{\partial}{\partial\varphi}+\rho\right),\label{Agd}\\
a_{1,2}^\dag& =\frac{1}{2}e^{\mp
i\varphi}\left(-\frac{\partial}{\partial\rho}\pm\frac{i}{\rho}\frac{\partial}{\partial\varphi}+\rho\right)\label{Agddag}
\end{align}
act on the complete eigenfunctions of the TDIHO as follows
\cite{COHEN}
\begin{eqnarray}
a_2\psi_{nm}=\sqrt{\frac{n+m}{2}}\psi_{n-1\;m-1},\hspace{.5in}a_2^\dag\psi_{nm}=\sqrt{\frac{n+m+2}{2}}\psi_{n+1\;m+1},\nonumber\\
a_1\psi_{nm}=\sqrt{\frac{n-m}{2}}\psi_{n-1\;m+1},\hspace{.5in}a_1^\dag\psi_{nm}=\sqrt{\frac{n-m+2}{2}}\psi_{n+1\;m-1}.\nonumber
\end{eqnarray}On the other hand, the constants of motion for this system,
expressed in polar coordinates, are \cite{JAUCH}
\begin{align}
O_\pm& =\mp
\frac{i}{2}e^{\pm2i\varphi}\left[\left(1\pm L_z\right)\left(\frac{1}{\rho}\frac{\partial}{\partial\rho}\mp\frac{L_z}{\rho^2}\right)+H_m\right],\label{O+-}\\
O_3&
=\frac{L_z}{2}=-\frac{i}{2}\frac{\partial}{\partial\varphi}\label{O3}
\end{align}
and satisfy the $su(2)$ Lie algebra
\begin{align}
[O_\pm,O_3]& =\mp O_\pm,\\
[O_+,O_-]& =2 O_3.\label{aloab2}
\end{align}
From these commutation relations, we prove that
\begin{equation}
O_\pm\psi_{nm}\propto\psi_{n\;m\pm2}.
\end{equation}

Also, it is straightforward to show that
\begin{equation}
O_+=-i a_2^\dag a_1,\hspace{0.5in}O_-=i a_1^\dag
a_2.\label{o+-}
\end{equation}

In order to find how the SUSY operators and constants of motion of
the TDIHO are related, we apply the operators (\ref{Agd}) and
(\ref{Agddag}) on the states defined in (\ref{psioab}). Thus, we
obtain
\begin{align}
a_{1,2}\psi_{nm}&
=\frac{1}{2}\rho^{-\frac{1}{2}}e^{i(m\pm1)\varphi}{\cal O}_{1,2}(m)u_{nm},\\
a_{1,2}^\dag\psi_{nm}&
=\frac{1}{2}\rho^{-\frac{1}{2}}e^{i(m\mp1)\varphi}{\cal O}_{1,2}^\dag(m)u_{nm},
\end{align}
where
\begin{align}
{\cal O}_{1,2}(m)&
=\frac{d}{d\rho}\mp\frac{m\pm\frac{1}{2}}{\rho}+\rho,\\
{\cal O}^\dag_{1,2}(m)&
=-\frac{d}{d\rho}\mp\frac{m\mp\frac{1}{2}}{\rho}+\rho.
\end{align}

From the expressions above, the following identities hold
\begin{align}
{\cal O}_1(m)& =-\sqrt{2}B_1^\dag,& {\cal
O}_1^\dag(m+1)& =-\sqrt{2}B_1,\\
{\cal O}_2(m)& =-\sqrt{2}B_2^\dag,& {\cal
O}_2^\dag(m-1)& =-\sqrt{2}B_2.
\end{align}

This shows that the SUSY operators are directly contained in the
operators $a_{1,2}$ and  $a_{1,2}^\dag$. Therefore, from the
expressions (\ref{o+-}), the radial spherical constants of
motion for the TDIHO are a product of radial SUSY operators.

\pagebreak
\thispagestyle{empty}
\begin{figure}
\centering \setlength{\unitlength}{0.25mm}
\begin{picture}(400,230)
\put(0,30){\vector(1,0){380}} \put(190,27.5){\vector(0,1){180}}
\multiput(40,27.5)(30,0){11}{\line(0,1){5}}
\multiput(187.5,60)(0,60){3}{\line(1,0){5}} \linethickness{0.5mm}
\multiput(180,30)(0,60){3}{\line(1,0){20}}
\multiput(150,60)(0,60){3}{\line(1,0){20}}
\multiput(120,90)(0,60){2}{\line(1,0){20}}
\multiput(90,120)(0,60){2}{\line(1,0){20}}
\multiput(60,150)(0,60){1}{\line(1,0){20}}
\multiput(30,180)(0,60){1}{\line(1,0){20}}
\multiput(210,60)(0,60){3}{\line(1,0){20}}
\multiput(240,90)(0,60){2}{\line(1,0){20}}
\multiput(270,120)(0,60){2}{\line(1,0){20}}
\multiput(300,150)(0,60){1}{\line(1,0){20}}
\multiput(330,180)(0,60){1}{\line(1,0){20}}
\put(175,60){\makebox(0,0){\scriptsize 1}}
\put(175,90){\makebox(0,0){\scriptsize 2}}
\put(175,120){\makebox(0,0){\scriptsize 3}}
\put(175,150){\makebox(0,0){\scriptsize 4}}
\put(175,180){\makebox(0,0){\scriptsize 5}}
\put(200,200){\makebox(0,0){$n$}}
\put(190,20){\makebox(0,0){\scriptsize 0}}
\put(159,20){\makebox(0,0){\scriptsize -1}}
\put(129,20){\makebox(0,0){\scriptsize -2}}
\put(99,20){\makebox(0,0){\scriptsize -3}}
\put(69,20){\makebox(0,0){\scriptsize -4}}
\put(39,20){\makebox(0,0){\scriptsize -5}}
\put(220,20){\makebox(0,0){\scriptsize 1}}
\put(250,20){\makebox(0,0){\scriptsize 2}}
\put(280,20){\makebox(0,0){\scriptsize 3}}
\put(310,20){\makebox(0,0){\scriptsize 4}}
\put(340,20){\makebox(0,0){\scriptsize 5}}
\put(370,20){\makebox(0,0){$m$}} \put(160,60){\vector(1,1){30}}
\put(160,120){\vector(-1,-1){30}} \put(40,180){\vector(1,-1){30}}
\put(130,150){\vector(-1,1){30}} \linethickness{.2mm}
\linethickness{.1mm}
\put(250,150){\vector(0,-1){60}}
\put(220,60){\vector(0,1){60}}
\put(250,155){\vector(1,0){60}} \put(340,185){\vector(-1,0){60}}
\put(165,80){\makebox(0,0){\scriptsize$a_2^\dag$}}
\put(140,111){\makebox(0,0){\scriptsize$a_2$}}
\put(280,162){\makebox(0,0){\scriptsize$O_+$}}
\put(310,192){\makebox(0,0){\scriptsize$O_-$}}
\put(125,168){\makebox(0,0){\scriptsize$a_1^\dag$}}
\put(65,167){\makebox(0,0){\scriptsize$a_1$}}
\put(210,100){\makebox(0,0){\scriptsize$D_+^n$}}
\put(240,130){\makebox(0,0){\scriptsize$D_-^n$}}
\end{picture}
\end{figure}

{\small Figure 1. Action of the operators $D_\pm^n$, $O_\pm$, $a_{1,2}$ and $a_{1,2}^\dag$ on the states of the TDIHO.}\\

\section{Two-dimensional hydrogen atom}
\subsection{Schr\"odinger factorization}
The Schr\"odinger equation for the two-dimensional hydrogen atom is
\begin{equation}
H_{h.a.}{\tilde\psi}(\rho,\varphi)\equiv\left[-\frac{1}{2}\left(\frac{\partial^2}{\partial\rho^2}+\frac{1}{\rho}\frac{\partial}{\partial\rho}+
\frac{1}{\rho^2}\frac{\partial^2}{\partial\varphi^2}\right)-\frac{1}{\rho}\right]{\tilde\psi}(\rho,\varphi)={\tilde E}{\tilde\psi}(\rho,\varphi).
\label{Hha}
\end{equation}
where we have set the electric charge equal to unity.

Since $[H_{h.a},L_z]=0$, the eigenfunctions are written as
\begin{equation}
{\tilde\psi}_{nm}(\rho,\varphi)={\tilde
R}_{nm}(\rho)e^{im\varphi}\equiv\rho^{-\frac{1}{2}}{\tilde
u}_{nm}(\rho)e^{im\varphi} \label{psiha}
\end{equation}
and the radial Hamiltonian for this system satisfies
\begin{equation}
{\tilde H}_m{\tilde
u}_{nm}\equiv-\frac{1}{2}\frac{d^2u}{d\rho^2}+\left[
-\frac{1}{\rho}+\frac{1}{2}\frac{m^2-\frac{1}{4}}{\rho^2}\right]{\tilde
u}_{nm}={\tilde E}_n{\tilde u}_{nm}. \label{habra}
\end{equation}
For bound states, we define $K^2_n=-\frac{1}{2{\tilde E}_n}$ and $\rho=K_n x$. Thus, equation
(\ref{habra}) can be rewritten as
\begin{equation}
\left( x^2\frac{d^2}{dx^2}+2K_n x-x^2\right){\tilde
u}_{nm}=\left(m^2-\frac{1}{4}\right){\tilde u}_{nm}. \label{hids}
\end{equation}
Following the procedure applied to the TDIHO, the Schr\"odinger
operators that change the principal quantum number $n$ are
\begin{equation}
T_\pm^n=\mp x\frac{d}{dx}+x-K_n,
\end{equation}
which satisfy
\begin{eqnarray}
(T_-^n-1)T_+^n
{\tilde u}_{nm}=\left[K_n(K_n+1)-\left(m^2-\frac{1}{4}\right)\right]{\tilde u}_{nm},\\
(T_+^n+1)T_-^n {\tilde
u}_{nm}=\left[K_n(K_n-1)-\left(m^2-\frac{1}{4}\right)\right]{\tilde
u}_{nm}
\end{eqnarray}
and their recurrence relations are given by
\begin{align}
T_+^n{\tilde u}_{nm}& \propto{\tilde u}_{n+1\;m},\\
T_-^n{\tilde u}_{nm}& \propto{\tilde u}_{n-1\;m}.
\end{align}

From (\ref{hids}), we define the operator
\begin{equation}
T_3=\frac{1}{2}\left(-x\frac{d^2}{dx^2}+x+\frac{m^2-\frac{1}{4}}{x}\right),
\label{T3}
\end{equation}
which allows us to generalize $T_{\pm}^n$ to
\begin{equation}
T_{\pm}=\mp x\frac{d}{dx}+ x-T_3.
\label{T+-}
\end{equation}
We show that these operators close the $so(2,1)$ Lie algebra
\begin{eqnarray}
[T_\pm,T_3]=\mp T_\pm,\\
\lbrack T_+,T_-\rbrack=-2T_3.
\end{eqnarray}

Notice that the non-compact algebras generators (Eqs. (\ref{D3}),
(\ref{D+-}), (\ref{T3}) and (\ref{T+-})) were introduced previously
in Ref. \cite{COOPER} for the case of the three-dimensional
problems. However, the origin of these operators was not investigated.

\pagebreak
\begin{figure}
\centering \setlength{\unitlength}{0.25mm}
\begin{picture}(400,230)
\put(0,10){\vector(1,0){380}} \put(190,7.5){\vector(0,1){200}}
\multiput(40,7.5)(30,0){11}{\line(0,1){5}} \linethickness{0.5mm}
\multiput(180,30)(0,30){6}{\line(1,0){20}}
\multiput(150,60)(0,30){5}{\line(1,0){20}}
\multiput(120,90)(0,30){4}{\line(1,0){20}}
\multiput(90,120)(0,30){3}{\line(1,0){20}}
\multiput(60,150)(0,30){2}{\line(1,0){20}}
\multiput(30,180)(0,30){1}{\line(1,0){20}}
\multiput(210,60)(0,30){5}{\line(1,0){20}}
\multiput(240,90)(0,30){4}{\line(1,0){20}}
\multiput(270,120)(0,30){3}{\line(1,0){20}}
\multiput(300,150)(0,30){2}{\line(1,0){20}}
\multiput(330,180)(0,30){1}{\line(1,0){20}}
\put(175,30){\makebox(0,0){\scriptsize 1}}
\put(175,60){\makebox(0,0){\scriptsize 2}}
\put(175,90){\makebox(0,0){\scriptsize 3}}
\put(175,120){\makebox(0,0){\scriptsize 4}}
\put(175,150){\makebox(0,0){\scriptsize 5}}
\put(175,180){\makebox(0,0){\scriptsize 6}}
\put(200,200){\makebox(0,0){$n$}}
\put(190,0){\makebox(0,0){\scriptsize 0}}
\put(159,0){\makebox(0,0){\scriptsize -1}}
\put(129,0){\makebox(0,0){\scriptsize -2}}
\put(99,0){\makebox(0,0){\scriptsize -3}}
\put(69,0){\makebox(0,0){\scriptsize -4}}
\put(39,0){\makebox(0,0){\scriptsize -5}}
\put(220,0){\makebox(0,0){\scriptsize 1}}
\put(250,0){\makebox(0,0){\scriptsize 2}}
\put(280,0){\makebox(0,0){\scriptsize 3}}
\put(310,0){\makebox(0,0){\scriptsize 4}}
\put(340,0){\makebox(0,0){\scriptsize 5}}
\put(235,75){\makebox(0,0){\scriptsize$T_+^n$}}
\put(145,75){\makebox(0,0){\scriptsize$T_-^n$}}
\linethickness{0.1mm}
\put(160,90){\vector(0,-1){30}}
\put(220,60){\vector(0,1){30}}
\put(370,0){\makebox(0,0){$m$}} \linethickness{0.2mm}
\put(220,125){\vector(1,0){30}} \put(310,185){\vector(-1,0){30}}
\put(160,125){\vector(-1,0){30}} \put(70,185){\vector(1,0){30}}
\put(240,135){\makebox(0,0){\scriptsize$F_1^\dag,\;F_2$}}
\put(290,195){\makebox(0,0){\scriptsize$F_1,\;F_2^\dag$}}
\put(140,135){\makebox(0,0){\scriptsize$G_-$}}
\put(90,195){\makebox(0,0){\scriptsize$G_+$}}
\end{picture}
\end{figure}
{\small Fig 2. Action of the operators $T_\pm^n$, $G_\pm$, $F_{1,2}$ and
$F_{1,2}^\dag$ on the states of the TDHA.}

\subsection{SUSY QM}
For the TDHA the SUSY operators which factorize the Hamiltonian
${\tilde H}_m$ are
\begin{align}
F_{1,2}&=\frac{1}{\sqrt{2}}\left[\frac{d}{d\rho}\pm\frac{m\pm\frac{1}{2}}{\rho}\mp\frac{
1}{m\pm\frac{1}{2}}\right],\label{F}\\
F_{1,2}^\dag&=\frac{1}{\sqrt{2}}\left[-\frac{d}{d\rho}\pm\frac{m\pm\frac{1}{2}}{\rho}\mp\frac{
1}{m\pm\frac{1}{2}}\right]\label{Fdag}
\end{align}
and the partners Hamiltonians are given by
\begin{eqnarray}
H_{1,2\;+}\equiv F_{1,2}F_{1,2}^\dag={\tilde H}_m+\frac{
1}{2(m\pm\frac{1}{2})^2},\label{Ha+}\nonumber\\
H_{1,2\;-}\equiv F_{1,2}^\dag F_{1,2}={\tilde H}_{m\pm1}+\frac{
1}{2(m\pm\frac{1}{2})^2}.\label{Ha-}\nonumber
\end{eqnarray}
It is straightforward  to show that, if ${\tilde u}_{nm}$ satisfies
(\ref{habra}), then
\begin{align}
F_{1}{\tilde u}_{nm}& \propto {\tilde u}_{n\;m-1},\label{f1}\\
F_{1}^\dag {\tilde u}_{nm}& \propto {\tilde u}_{n\;m+1}\label{f1dag}
\end{align}
and
\begin{align}
F_{2}{\tilde u}_{nm}& \propto {\tilde u}_{n\;m+1},\label{f2}\\
F_{2}^\dag {\tilde u}_{nm}& \propto {\tilde
u}_{n\;m-1}.\label{f2dag}
\end{align}

The action of these operators on the radial function ${\tilde
u}_{nm}$ is shown in figure 2. From expressions above, we conclude
that the action of the operators $F_1$ and $F_1^\dag$ on ${\tilde
u}_{nm}$ is identical to that of the operators $F_2^\dag$ and $F_2$,
respectively. Thus, a pair of these is redundant and can be left
out, say $F_2$ and its conjugate.

\subsection{SUSY QM and constants of motion}
In polar coordinates, the constants of motion for the TDHA given in
\cite{JAUCH} can be written as
\begin{align}
G_\pm& =\frac{1}{\sqrt{2|\tilde E|}}\rho e^{\pm
i\varphi}\Big\lbrace\left(\frac{1}{2}\pm L_z\right)\left(\frac{1}{\rho}
\frac{\partial}{\partial\rho}\mp\frac{L_z}{\rho^2}\right)+\frac{1}{\rho}\Big\rbrace,\\
G_3& =L_z=-i\frac{\partial}{\partial \varphi}
\end{align}
and satisfy
\begin{align}
[G_+,G_-]& =2 G_3,\\
[G_\pm,G_3]& =\mp G_\pm.\label{rot}
\end{align}
Thus, $G_\pm$ and $L_z$ define the $su(2)$ Lie algebra. If we
consider that the operators $L_z$ and $H_{h.a.}$ have simultaneous
eigenfunctions, from equation (\ref{rot}) we show that
\begin{equation}
G_\pm\psi_{nm}\propto\psi_{n\;m\pm1}. \label{G+-}
\end{equation}

In order to find how the constants of motion of the TDHA and SUSY
operators are related, we apply  the operators $G_\pm$ on the states
given in (\ref{psiha}). Therefore, we obtain
\begin{equation}
G_\pm\psi_{nm}=-\frac{1}{\sqrt{2|\tilde E|}}\left(m\pm\frac{1}{2}\right)\rho^{-\frac{1}{2}}e^{i(m\pm1)\varphi}g_\pm^m
u_{nm},
\end{equation}
where we have defined the radial operators
\begin{equation}
g_\pm^m=\mp\frac{d}{d\rho}+\frac{m\pm\frac{1}{2}}{\rho}-\frac{1}{m\pm\frac{1}{2}}.
\label{g}
\end{equation}
Equations (\ref{F}), (\ref{Fdag}) and (\ref{g}) lead us to
\begin{equation}
F_1=\frac{1}{\sqrt{2}}g^{m+1}_-,
\end{equation}
\begin{equation}
F^\dag_1=\frac{1}{\sqrt{2}}g^m_+.
\end{equation}
Therefore, unlike to the TDIHO, the SUSY operators are directly
contained in the spherical constants of motion.

\section{The $2D$ potential $A\rho^{2\zeta-2}-B\rho^{\zeta-2}$: zero-energy eigensubspace}
The purpose of this section is to find the compact and non-compact
algebras of a general potential which includes those of the TDIHO
and TDHA as particular cases. We consider the following
two-dimensional Hamiltonian \cite{KHARE1,KHARE2}
\begin{equation}
{\cal
H}\equiv-\frac{1}{2}\left(\frac{\partial^2}{\partial\rho^2}+\frac{1}{\rho}\frac{\partial}{\partial\rho}+
\frac{1}{\rho^2}\frac{\partial^2}{\partial\varphi^2}\right)+A\rho^{2\zeta-2}-B\rho^{\zeta-2},
\label{Hgen}
\end{equation}
where $A,\;B$ $>0$, and  $\zeta$ is a positive rational number. It
has been shown that the zero-energy level for this operator
possesses accidental degeneracy \cite{KHARE1,KHARE2,VINET}. Let
$\Psi_0(\rho,\varphi)$ one state of the zero-energy eigensubspace,
this is
\begin{equation}
{\cal H}\Psi_0=0. \label{Hcero}
\end{equation}
If we define the operator
\begin{equation}
H\equiv
\rho^{2-\zeta}\left[-\frac{1}{2}\left(\frac{\partial^2}{\partial\rho^2}+\frac{1}{\rho}\frac{\partial}{\partial\rho}+
\frac{1}{\rho^2}\frac{\partial^2}{\partial\varphi^2}\right)+A\rho^{2\zeta-2}\right],
\label{Hhat}
\end{equation}
then from Eq.(\ref{Hcero}), it follows that
\begin{equation}
H\Psi_0(\rho,\varphi)=B\Psi_0(\rho,\varphi). \label{Hhat01}
\end{equation}
We note that Eq. (\ref{Hhat01}) encompasses the Sch\"rodinger
equation for the $2D$ isotropic harmonic oscillator ($\zeta=2$) and the
$2D$ hydrogen atom ($\zeta=1$).

By setting $\rho^{\zeta}=\frac{\zeta}{2\sqrt{2A}}y^2$, equation
(\ref{Hhat01}) is transformed to
\begin{equation}
{\tilde
H}\Psi_0(y,\theta)\equiv\left[-\frac{1}{2}\left(\frac{\partial^2}{\partial
y^2}+\frac{1}{y}\frac{\partial}{\partial
y}+\frac{1}{y^2}\frac{\partial^2}{\partial\theta^2}\right)+\frac{y^2}{2}\right]\Psi_0(y,\theta)=\Lambda\Psi_0(y,\theta),
\label{Hy}
\end{equation}
where $\theta=\frac{\zeta}{2}\varphi$, $\Lambda=\frac{2B}{\zeta\sqrt{2A}}$ and $\Psi_0(y,\theta)=R(y)e^{im\theta}$. Eq. (\ref{Hy})
corresponds to the Schr\"odinger equation of the TDIHO in polar
coordinates $(y,\theta)$. This result will allow us to find the generators of the compact and
non-compact symmetries of the operator $H$.

\subsection{Constants of motion and spectrum generating algebras}
Considering the last result, the constants of motion of the
Hamiltonian $\tilde H$ are obtained from Eqs. (\ref{O+-}) and
(\ref{O3}), and are given by
\begin{align}
F_\pm& =\mp ie^{\pm2i\theta}\left[\left(1\pm
L_{z'}\right)\left(\frac{1}{y}\frac{\partial}{\partial
y}\mp\frac{L_{z'}}{y^2}\right)+{\tilde
H}\right],\label{F+-}\\
F_3&
=-\frac{i}{2}\frac{\partial}{\partial\theta}\equiv\frac{L_{z'}}{2}\label{F3},
\end{align}
whereas, from Eqs. (\ref{D3}) and (\ref{D+-}), the Schr\"odinger
operators are
\begin{align} K_\pm& =\frac{1}{2}\left(\mp
y\frac{d}{dy}+y^2-2K_3\mp\frac{1}{2}\right),\\
K_3& =\frac{\tilde H}{2}.
\end{align}

By performing the change of variables
$(y,\theta)\rightarrow(\rho,\varphi$), the constants of motion are
\begin{align}
\Theta_\pm& =\mp\frac{i}{\zeta\sqrt{2A}}\rho^{2-\zeta}e^{\pm
i\zeta\varphi}\left\lbrace\left(\frac{\zeta}{2}\pm L_z\right)
\left(\frac{1}{\rho}\frac{\partial}{\partial\rho}\mp\frac{L_z}{\rho^2}\right)+\rho^{\zeta-2}H\right\rbrace,\label{genc1}\\
\Theta_3& =\frac{1}{\zeta}L_z \label{genc2}
\end{align}
and the Schr\"odinger operators are
\begin{align}
\Delta_3& =\frac{1}{2\zeta\sqrt{2A}}H,\label{gennc1}\\
\Delta_\pm&
=\frac{1}{2}\left(\mp\frac{2}{\zeta}\rho\frac{\partial}{\partial\rho}+\frac{2\sqrt{2A}}{\zeta}\rho^\zeta-2\Delta_3\mp1\right).\label{gennc2}
\end{align}
Notice that if $R(\rho)\equiv\rho^{-\frac{1}{2}}U(\rho)$ then, for $\zeta=1,2$ the operators above are reduced to the
compact and the non-compact algebra generators for the $2D$ hydrogen
atom and the $2D$ harmonic oscillator, respectively. On the other hand,
by considering that $\Theta_{\pm}=\Theta_1\pm i \Theta_2$,  we
obtain the operators which are essentially the symmetry generators
reported in Ref. \cite{VINET}.

We show that the operators $\Theta_\pm$ and $\Theta_3$ satisfy the
$su(2)$ Lie algebra
\begin{align}
[\Theta_+,\Theta_-]& =2\Theta_3,\\
[\Theta_\pm,\Theta_3]& =\mp\Theta_\pm.
\end{align}
and the operators $\Delta_\pm$ and  $\Delta_3$ close the $so(2,1)$
spectrum generating algebra
\begin{align}
[\Delta_+,\Delta_-]& =-2\Delta_3,\\
[\Delta_\pm,\Delta_3]& =\mp\Delta_\pm.
\end{align}

By transforming the operator $H$ to the extensively studied
Hamiltonian of the TDIHO, we were capable to find its spherical
constants of motion and the generators of the non-compact algebra.

\section{Concluding remarks}

The factorization techniques have been used to obtain algebraically
the wave functions and the energy spectrum of many systems. In this
paper, we showed in a systematic and clear way how factorization
methods and hidden and dynamical symmetries are related. By means of
the Schr\"odinger factorization we constructed the operators which
close the $so(2,1)$ spectrum generating algebra for the TDIHO and
the TDHA. We showed the link between the SUSY operators and the
symmetry generators of the $su(2)$ algebra for these systems.

The results concerning to the zero-energy eigensubspace for the $2D$
potential $A\rho^{2\zeta-2}-B\rho^{\zeta-2}$ were obtained by
transforming (\ref{Hhat01}) to the equation of the TDIHO. Thus, we found the origin of the hidden and
dynamical symmetry generators and showed that these operators close
the $su(2)$ and $so(2,1)$ Lie algebra, respectively.  We note that
our non-compact generators and those given in \cite{VINET1} are
different realizations of the algebra $so(2,1)$. Also, our procedure
clarifies the restrictions on the quantum numbers, the
eigenfunctions and the origin of the symmetries generators reported
in \cite{VINET1}.

Although the compact and non-compact algebra generators are
difficult to find, in this work we have shown that factorization
methods provide the explicit form of these operators. This seems to
be a first approach to a systematic method to find  hidden and
dynamical symmetries. Thus, we think that the determination of closed Lie algebras of a Hamiltonian is closer to science than art, contrary to the opinion expressed in \cite{MOSHI2}.

\section*{Acknowledgments}

This work was partially supported by  COFAA-IPN, EDI-IPN, SIP-IPN
projects numbers 20070717 and 20071135.

\end{document}